# Symbolic Manipulators Affect Mathematical Mindsets

Thomas J. Bing and Edward F. Redish
Department of Physics
University of Maryland, College Park, MD 20742-4111, USA

**Abstract**

Symbolic calculators like Mathematica are becoming more commonplace among upper level physics students. The presence of such a powerful calculator can couple strongly to the type of mathematical reasoning students employ. It does not merely offer a convenient way to perform the computations students would have otherwise wanted to do by hand. This paper presents examples from the work of upper level physics majors where Mathematica plays an active role in focusing and sustaining their thought around calculation. These students still engage in powerful mathematical reasoning while they calculate but struggle because of the narrowed breadth of their thinking. Their reasoning is drawn into local attractors where they look to calculation schemes to resolve questions instead of, for example, mapping the mathematics to the physical system at hand. We model Mathematica's influence as an integral part of the constant feedback that occurs in how students frame, and hence focus, their work.

01.40.Fk, 01.30.lb

## I. Introduction

Recent advances in computers and programming have given today's physics students a new tool. Personal computer programs such as Mathematica, Maple, and even handheld calculators are now widely capable of symbolic manipulation. Whereas calculators were once limited to numeric operations like evaluating the cube root of forty-two, they can now expand $(x+3)^3$ to $x^3 + 9x^2 + 27x + 27$, evaluate $\int_{-\infty}^{\infty} \frac{\cos(x)}{1+x^2} dx$ as $\frac{\pi}{e}$, and solve $\frac{dy}{dx} = -3y$ as $y(x) = Ce^{-3x}$.

Automated calculation, even when it is strictly numerical instead of symbolic, makes many teachers wary. Almost all physics teachers have anecdotal stories of watching students reach for a calculator to do simple operations like halving a number or multiplying by one hundred. Most have also watched students make obvious errors as they pushed calculator buttons. Teachers worry that these students are neither using nor developing a feel, an instinct, for numbers. Mathematics education researchers have been similarly concerned with this vital sense in students. "Number sense" has at least partially converged to a certain set of meanings in the math education literature, including flexible computing strategies for written and calculator-aided computation, understanding of equivalent representations, and use of equivalent expressions.[1]

With the expansion of symbolic manipulation capabilities into the teaching of advanced physics, we can raise an analogous question: Does being fluent with a symbolic manipulator damage the advanced physics student's intuition for and ability to make sense of complex mathematics in physics?

In this paper we begin to address this question by offering two examples from upper level physics majors as they use Mathematica to solve problems in an upper division course in quantum mechanics. Mathematica's presence contributes to the students' difficulties in both cases, but their difficulties do not stem from a stunted or disengaged mathematical intuition. The students show admirable flexibility and creativity as they try different calculation strategies and representational forms. Rather, the difficulties associated with Mathematica use appears to arise from more subtle issues. They arise from a local coherence in their thinking that leads them to focus on computational aspects of the problem while suppressing the connection with the physics and with extended mathematical meanings. We analyze these observations in terms of a theoretical framework based on studies of selective attention and framing in the cognitive and behavioral sciences.

## II. Computational Tools and Mathematical Intuition

In considering the student use of symbolic manipulators in advanced physics, it is appropriate to put our considerations in perspective of the use of computational tools at other levels. University instructors often feel that the calculator has done damage to the growth or use of the "number sense" in students. Is this so?

This is a broad question and is complicated by the fact that "number sense" is difficult to define exactly. There is a very large collection of studies with elementary through high school students (see Hembree and Dessart[2][3] or Dunham[4] for reviews) that mostly suggest numeric calculators help students develop algorithmic computation and problem solving skills. Better rote computation skills, however, do not necessarily imply a better number sense.[5] Still, the relatively few studies that explicitly address the effects of cal-culators on students' number sense tend to indicate calculators help.[6,7] A possible explanation for this phenomenon would be to conceptualize number sense as something that evolves out of one's interaction with a conceptual environment.[8] The more a student thinks, works, even plays around with mathematics, the more their intuition, their number sense, evolves. By analogy, the longer you live and navigate in a city, the better your sense of direction becomes. Calculators can help streamline this playing around in mathematics, providing quick feedback that can accelerate the development of number sense.

Once students move on to algebra, number sense is extended to include symbol sense. Symbol sense includes healthy intuitions about when introducing symbols can be useful, what passes for proper symbol manipulations, and how symbolic arguments can be general methods of proof.[9,10] Relatively few studies with symbolic manipulation calculators have explicitly addressed their effects on symbol sense, but some positive correlations exist.[11] Speaking more broadly, symbolic manipulator use tends to correlate to both better conceptual understanding and better manual calculation skills,[12,13] just as the numerical calculator studies indicate.

These studies suggest that appropriate instruction using the calculator at a pre-college level may help students develop a sound sense of number and symbol. Whether most of our students have received such appropriate instruction and have such a sense remains to be explored.

In the case of the use of topics such as algebra and calculus in advanced physics, we are concerned with something more than a sense of number and symbol. We want students to develop a "sense of the mathematics" – an intuition for the structure of complex mathematical expressions that allows them to interpret and unpack these expressions, providing a capability for transforming equations and quickly recognizing errors. We refer to this extension of the number and symbol senses to more complex mathematics as *math sense*. In the rest of this paper we explore in depth two examples of students working together to solve authentic physics problems using Mathematica.

### III. Two Examples

The data set from which this study is taken consists of videos of students working in groups on problems in advanced physics. The problems are authentic homework assignments for which the students receive class credit. We have taken approximately one hundred hours of such videos in upper-division physics-major classes that include intermediate mathematical physics, electromagnetism, and quantum mechanics. Of these, approximately 10% include student use of Mathematica or other symbolic calculator. These examples provide the clearest representatives of phenomena that we have seen many times.

### Example 1: The Feynman-Hellmann Theorem

As one example of Mathematica's influence on student behavior, we have a video of two students doing their homework in a second semester undergraduate quantum mechanics class. They are working on problem 6.32, part b, in D.J. Griffiths's text.[14] It asks them to use the Feynman-Hellmann theorem, $\frac{\partial E_n}{\partial \lambda} = \langle \psi_n | \frac{\partial H}{\partial \lambda} | \psi_n \rangle$, for the one-dimensional harmonic oscillator with the Hamiltonian $H = \frac{p^2}{2m} + \frac{1}{2} m\omega^2 x^2$. They are asked to set $\lambda = \omega$ to obtain a formula for the expectation value of *V*, the potential energy. The video shows the students engaged in a fifteen-minute effort to program Mathematica to explicitly calculate the expectation value $\langle \psi_n | \frac{\partial H}{\partial \omega} | \psi_n \rangle$.

Their work is impressive. They identify the Hamiltonian, navigate a complicated general expression for the stationary states of the oscillator, implement Mathematica's predefined Hermite polynomial function, and neatly package all this information into a single line of code that can calculate the expectation value of $\frac{\partial H}{\partial \omega}$. When they hit one of several snags along the way, their error checking and debugging are quick and efficient. To illustrate, consider the following continuous 90-second chunk of transcript from the middle of this episode.

1. Student A: Umm, Hermite polynomials are all real, right? They're happy? Are there "i"s in the Hermite polynomials?
2. Student B: Let me check. Remember they had the first set of them way back here.
3. SA: Mmm-hmm, they're all real.
4. SB: Yeah, they're all real.
5. SA: All right, so they're just psi squared.
6. SB: Oh, there's, one moment, OK.
7. SA: Psi squared, d-omega-H, comma x, comma
8. SB: Umm, are those all the different bits? Where's your e to the negative xi squared?
9. SA: It's inside.
10. SB: It is? OK. OK good.
11. SA: x comma, oh , just x—er well, minus infinity to infinity, right?
12. SB: Right.
13. SA: Minus escape-n, comma, escape-n.
14. SB: That's what you got.
15. SA: Yeah, 'cause it didn't do it.
16. SB: No it didn't, and it got something. Let's see, negative n, m, didn't even get the same thing I did. It's different.
17. SA: Hermite n of x gives a Hermite polynomial h, n of x.
18. SB: We got h, n of x?

19. SA: What the heck? n is n, oh, 'cause it doesn't know what n is.
20. SB: We don't want this for any n. You have to say what n is?
21. SA: Well, you can't integrate the Hermite polynomial of x without putting in what the Hermite polynomial is.

It takes a formidable math sense to accomplish these actions, one that goes well beyond the much simpler examples from grade school mathematics on which most of the number sense or symbol sense literature focuses. There are hardly any actual numbers in sight, only variables and constants that stand in for them. Complicated functions, Hermite polynomials and $\psi_n$, have to be seen as mathematical objects[15, 16] in and of themselves to be unpacked and operated upon, as in lines 1 to 5. Mathematica has its own protocol for using Hermite polynomials, referred to in lines 17 and 18, that the students have found and interpreted. They have also managed to organize the calculation in an efficient and aesthetic way, hiding some of the details behind user-defined symbols in the Mathematica code. S2 asks about one such move in line 8. After this snippet ends, the students even set up an array in Mathematica to perform ten of these expectation value calculations, one for each of the first ten stationary states, at once.

Their debugging is efficient as well. Lines 15 to 17 show the students reacting to Mathematica's evaluation of their first coding. The program has balked at the Hermite polynomial function call. S1 quickly interprets the error in lines 19 and 21. They have not indicated which specific Hermite polynomial Mathematica should use. Such a quick debugging demonstrates S1's engaged grasp of the Hermite polynomials. They are an articulated set of specific mathematical objects to him, not merely some nebulous symbol upon which to operate.

This transcript, and the larger fifteen-minute episode that surrounds it, is quite a display of flexible computation and representation, two hallmarks of math sense. In their approach to this activity, we say that the students have a mindset that focuses on *drilling down into a calculation*. In this state, students pay attention to the calculational details, look for ways to achieve a result, unpack mathematical structures, and manipulate expressions within the problem that they have identified.

Absent from all this work, however, is any discussion of how they plan to connect their calculation's result to <V> as the question requires. Their thinking was drawn into this Mathematica calculation, which sustained itself for fifteen straight minutes even when difficulty arose. Making the calculation work became a goal in and of itself, irrespective of the original homework question. Their excellent calculation eventually yields the truism $\hbar\left(n + \frac{1}{2}\right) = \hbar\left(n + \frac{1}{2}\right)$ from the Feynman-Hellmann theorem, but then they are stuck.

The never step back to notice that $\frac{\partial H}{\partial \omega}$ is proportional to $x^2$, as is *V*. Simply shuffling a few constants around in the Feynman-Hellmann theorem can yield an expression for <V>. No explicit calculation of expectation values is required.

This is not to say that thinking about how both $\left\langle\frac{\partial H}{\partial \omega}\right\rangle$ and <V> are proportional to <$x^2$> is not also an application of math sense. It is, however, an application driven by a search for a different kind of mathematical justification. When these students were programming Mathematica to compute the expectation value of $\frac{\partial H}{\partial \omega}$, they were focusing on how convincing mathematical arguments are procedurally correct. Technically correct calculation should lead to a trustable result. Their number sense is projected along this computation axis and manifests itself as the flexible calculation and representation strategies seen in their work.

Noticing that $\left\langle\frac{\partial H}{\partial \omega}\right\rangle$ and <V> are both proportional to <$x^2$> focuses on a different aspect of mathematical justification. Instead of being concerned with drilling down into a detailed calculation, it entails packaging parts of an expression together and seeing how the various packages relate to one another.

If the previous computation mindset was "drilling down" then this present mindset is more of a "moving across". We will call this mindset *mathematical chunking*. Chunking is also an important type of mathematical justification. Mathematical systems involve many parts, and understanding how each part interacts and relates to the other parts is essential for comprehending the system as a whole. Whereas the earlier computation mindset brought out certain facets of a student's math sense, the mathematical chunking mindset highlights other aspects of math sense, such as proportionality and functional dependence.

In this example, Mathematica seems to have facilitated the students entering and sustaining a calculationally-focused mindset, ignoring broader and more direct mathematical approaches to the problem. This is not to say that the students would not have chosen a calculational mindset without Mathematica. Indeed, we have seen many students doing this sort of thing at many levels. But having Mathematica seems to remove a barrier to entering a computational mode that is explicitly illustrated in the next example.

### Example 2: An Expectation Value

Our second example comes from a video recording of six junior and senior physics majors meeting to work on their homework for a second semester undergraduate quantum mechanics class. They are working on Problem 5.6 in D.J. Griffiths's text.[14] The problem asks them to calculate <$(x_1 - x_2)^2$> for two particles in arbitrary stationary states of a one-dimensional infinite well, where $x_1$ is the coordinate of the first particle and $x_2$ is the coordinate of the second. Three successive parts of the problem ask them to assume the particles are distinguishable, identical bosons, and identical fermions. In the course of this calculation, the students realize they need to evaluate $\int x_1^2 |\psi_n(x_1)|^2 dx_1$.

This notational shorthand, which doesn't specify the limits of integration, is taken from the hints the text gives in the pages preceding this problem. The transcript begins with a student in the group explicitly mistaking the limits of integration to be from negative infinity to positive infinity instead of just over the width of the well. They are thus led to try to evaluate $\frac{2}{L}\int_{-\infty}^{\infty} x^2 \sin^2\left(\frac{n\pi x}{L}\right) dx$.

1. S1: The integral is from negative infinity to infinity, right?
2. S2: Yeah.
3. S1: So we have x squared *(types in Mathematica)*
   …one minute later…
4. S1: It's telling me it doesn't converge. What if I tried *(sets Mathematica aside, begins trying to integrate by parts with pencil and paper)*
5. S3: So what's the integral equal to?
6. S1: It wasn't happy, so let me just try something else.
7. S3: Oh, we got undefined?
8. S1: It said it didn't converge.

S1 is our main focus. She is one of the top students in her class and graduated with honors and significant research experience. Our analysis of her thinking centers around five times when she explicitly hits a roadblock in her work during the seven-minute stretch from which these transcript chunks are drawn. By *hits a roadblock*, we mean that her current line of thinking has either come to a result that does not satisfy her or that has become too complicated to justify continuing. The most important aspect shared by the roadblocks S1 encounters is that they all necessitate her picking a new approach.

S1 encounters five roadblocks and makes five choices about what is appropriate to try next. All of her choices result in strategies aimed at producing a technically correct calculation except for one ambiguous case at the end. Mathematica is an integral part of her thinking during each of the events we observe.

S1 encounters the first of these roadblocks above in line 4. She mistakenly sets the limits of integration in line 1, and Mathematica correctly informs her that $\frac{2}{L}\int_{-\infty}^{\infty} x^2 \sin^2\left(\frac{n\pi x}{L}\right) dx$ diverges. Faced with this unexpected result, S1 now faces a choice of how to proceed. She chooses to try evaluating the integral by hand. This choice may or may not have been a result of conscious reflection. Note that whether S1 consciously thought of an alternative way to continue and then suppressed it in favor of integrating by parts manually is not directly relevant. What is relevant is the fact that her antidote to the failed Mathematica calculation is another form of calculation.

S1 started by trying to answer the question "What is the value of $\frac{2}{L}\int_{-\infty}^{\infty} x^2 \sin^2\left(\frac{n\pi x}{L}\right) dx$?" The initial Mathematica computation was aiming to produce and justify a result by means of a technically correct calculation. She keeps her search for justification in the calculational realm even though the roadblock has now transformed the original question with the refinement "Does $\frac{2}{L}\int_{-\infty}^{\infty} x^2 \sin^2\left(\frac{n\pi x}{L}\right) dx$ really diverge?" A calculation strategy is by no means the only type of justification to use in answering this question. Mathematical chunking would work well. One could, for example, sketch a graph of the integrand or consider the character of its parts. A squared sine function is neither negative nor does it tend asymptotically to zero, and $x^2$ certainly tends to infinity as $x$ approaches positive and negative infinity. That integral must therefore blow up. S1, however, keeps her search for proof in the calculation realm.

The next strip of transcript picks up about ten seconds after the end of the previous strip.

9. S1: I mean, this is an integral that's quite do-able by *(brings back computer with Mathematica)*
10. S3: trig substitution
11. S1: by parts
12. S3: oh, by parts
13. S2: Yeah.
14. S1: So *(starts typing again)*
15. S4: Can you break it up into different parts and then do it on a TI-89? That's what I usually do, a combination by hand, by calculator.
16. S3: Well, integrate it indefinitely and plug in.
17. S5: Are you not substituting a value in for n and L, or are you?
18. S1: Umm, no, but I just tried doing x-squared, sine of x squared, and it's not happy.

S1 implicitly encounters her second roadblock in line 9. She had been trying to manually integrate by parts with pencil and paper but decided such a calculation would be too involved to reasonably continue. Again a choice of new direction confronts S1, and she again opts for another calculational approach centered on using the computer. She reaches for Mathematica again and tries evaluating a simpler form of the integral, $\int_{-\infty}^{\infty} x^2 \sin^2 x\, dx$, as she reports in line 18.

This incident is an example of how having Mathematica as a tool can "open channels" to calculational approaches that might not have been chosen had it not been available. It costs S1 a very small effort investment to try evaluating this slightly different integral. Mathematica lowers the potential barrier to the evaluation of $\int_{-\infty}^{\infty} x^2 \sin^2 x \, dx$, allowing S1 to explore the problem space more freely. The downside, as this example will illustrate, is this calculation enabling can make it that much easier to get stuck in a calculation mindset.

Also noteworthy is how the local tendency to solve this dilemma solely by further calculation spreads through the group. The rest of the group sees S1 reach for Mathematica a second time in line 9 and infers she needs help with the divergent result. Three other students offer potential solutions, all of which are calculation strategies. S4 suggests a hybrid approach in line 15. Do the potentially complicated work of rearranging $\int u\,dv$ into $uv - \int v\,du$ by hand, and only then call on the computer to work on the simpler integrals. Line 16 has S3 suggesting Mathematica might be having trouble evaluating the antiderivative at the positive and negative infinity limits. Try just letting Mathematica find the indefinite integral of $x^2 \sin^2\left(\frac{n\pi x}{L}\right)$ and then plug in the limits by hand. S5 offers, in line 17, that maybe Mathematica is being confused by an undefined parameter.

All of these suggestions, in addition to the one S1 has tried in line 18, reflect a developed, engaged math sense. They treat the calculation at hand as a malleable thing, as something that can be rearranged, simplified, and executed in different ways. The explicit representation of the integral is changed as the students work.

The suggestions of all these students reflect a sophisticated perception that Mathematica is a fallible tool whose precise usage can be deconstructed and tailored to suit the situation at hand. Interpreting the activity as one of calculation does not imply naivety or unsophisticated reasoning. Their difficulty, like the students in the Feynman-Hellmann example, does not stem from a math sense muted by Mathematica. It comes from the relative narrowness of their search. They are trying to resolve a calculational difficulty with more calculation instead of thinking about the integral itself or asking how the integral they are trying to calculate aligns with the physical situation at hand. These alternate framings would bring out different facets of their math sense.

With the failure of her simpler Mathematica calculation in line 18, S1 encounters her third roadblock. She again elects to try more calculation to resolve it and proceeds to follow some combination of S3 and S4's suggestions. She types some more into Mathematica, produces the antiderivative of the integrand, and then spends nearly a minute copying the antiderivative from her computer screen onto her paper. S1 is looking at this antiderivative when she next begins speaking.

19. S1: I can see why it says that doesn't converge.
20. S2: Yeah, but I know it...we've done it.
21. S6: We're like, but I know it does.
22. S2: We've done that integral so many times.
23. S1: Find me one, cause see, this (indefinite integral)
24. S2: Yeah.
25. S1: Is equal to that (antiderivative), and so you know there's a whole number of places where it'll shoot to infinity.
26. S2: Like, how else do we find the expectation value of x-squared?
27. S1: Yeah.
28. S2: Like, I know we've done it for the infinite square well. *(S1 starts paging back through textbook)*

S1 succeeded in drilling down into the calculation Mathematica does when it tries to evaluate $\frac{2}{L}\int_{-\infty}^{\infty} x^2 \sin^2\left(\frac{n\pi x}{L}\right) dx$. In line 19, she shows S2 the various places in the antiderivative where plugging in the infinity limits leads to an infinite result. When both S2 and S6 respond to her work by asserting the result must be finite in lines 20-22, S1 faces her fourth roadblock. She elects to trust her result. No refocusing occurs as she counts her technically correct calculation as sufficient justification. In line 23, she challenges S2 to find an example of the "so many times" they've allegedly done this integral and proceeds to summarize her calculation for him.

S1 faces a fifth roadblock when S2 refines his finite-value assertion in lines 26 and 28. This specific integral has occurred much earlier in their quantum mechanics coursework when they were simply calculating $<x^2>$ for a single particle in an infinite well. That result was not infinite. Her response to this final roadblock is ambiguous. She does not say anything more but begins paging back through her textbook. By cognitive inertia, one might expect she is looking back to the book's original infinite well treatment, searching for an explicit calculation of $<x^2>$. There is no evidence to confirm or deny this assumption, however, because thirty seconds later S3 speaks.

29. S3: Hey, it's not negative infinity to infinity.
30. S1: What is it?
31. S3: Is it? Well, we just have to integrate it over the square well, 'cause it's the infinite square well.
32. S2: Oh yeah, so it's zero to [L].
33. S1: *(chuckling)* You're right.
34. S3: Yeah, that's why it's not working.
35. S1: Well, is it zero to [L] or negative [L] to [L]?
36. S2: Uhh, it's defined in [chapter] 2.2 as zero to [L].
37. S3: So yeah, that would be why we're [dumb]. *(laughs)*

38. S5: Oh. We're awesome.
39. S3: Yeah, none of us know how to do a square well anymore. *(laughs)*
40. S6: We really know what we're doing.
41. S4: What are you guys talking about?

In line 29, S3 tracks down the cause of the group's difficulties. He interpreted the task differently, looking for a different type of justification for his mathematics. Instead of looking towards more and more detailed calculation as S1 and the group have been doing, S3 has now looked at the fit of the mathematics they are using with the physical system under consideration. The negative to positive infinity limits of integration do not match the finite span of the infinite well. S3 had shown inklings of this shift towards thinking about what the math was modeling about ninety seconds earlier when, in an unquoted part of the transcript, he had asked which of the group's derivations were meant to correspond to distinguishable particles, fermions, and bosons. When the camera panned over to S3 directly after his pivotal comment in line 29, his calculator was not obviously positioned around him.

How do we justify calling S3's new mode of thought a significant shift? Most importantly, the students give clues in their speech that indicate they feel S3 has done a different type of thinking than they have been doing. S1 chuckles in line 33 as she acknowledges S3's answer. This laughter could indicate several things about S1's thought. Perhaps, like many other instances of laughter, it indicates surprise or a violation of an expected action. S1 was expecting more and different types of calculation, and S3's new contribution fell outside of that expectation. Perhaps it is an embarrassed laugh. S1 is maybe a little ashamed of how she was temporarily blinded to this relatively straightforward solution. In either case, her laughter indicates that she feels she has been doing a different type of thinking than was needed. S3 also laughs as he pokes fun at himself and the group in lines 37 and 39.

Two other students react with sarcasm, a close cousin of the S1 and S3's laughter. S5 and S6 sarcastically compliment themselves and the group. This sarcasm, regardless of whether it is more indicative of embarrassment or exasperation, indicates that S5 and S6 are also aware of the temporary blindness that has affected the group.

S4 provides a final piece of evidence that the group is itself aware of a shift in their thinking. He has been reading the text silently for most of this last snippet, but he asks the rest of the group what has just happened in line 41. His question suggests he has noticed the sudden change in the conversation's composition, the laughter and sarcasm described earlier. While he has missed the content of the shift, the change in tone that accompanies the other students' reframing still communicates "something different is going on here" to S4. The new tone communicates so strongly that S4 is compelled to explicitly ask what just happened.

The focus at the end of this transcript illustrates a different mindset from calculation or mathematical chunking, *physical mapping*. This third mindset is especially important for math use in physics. It is the examination of the interplay between the physical system at hand and the mathematics used to model it. This mindset highlights how mathematics in physics class is only valid insofar as it reflects the physical system under study. This mindset highlights still different components of a student's math sense, those focusing especially on the physical meaning behind numbers and their operations.

### IV. Discussion

In these two examples, we have observed students working in what we have referred to using the non-technical term "mindset." This term indicates that the student, or group of students, is temporarily focusing on a limited subset of their available tools and skills. Of these three mindsets, calculation, chunking, and physical mapping, Mathematica couples strongly to the calculation one.[17]

In the first example, the students spent fifteen minutes calculating $\left\langle \frac{\partial H}{\partial \omega} \right\rangle$, persisting even through difficulties. Their work was neither naïve nor silly; their math sense was engaged. However, this calculation mindset, influenced by Mathematica's presence, highlighted certain aspects of their math sense at the expense of others.

The second example illustrates how Mathematica plays a role in providing feedback that encourages the students to remain in a calculational mindset. Again, their trouble doesn't come from lacking math sense but rather stems from applying that math sense narrowly towards computational issues. Mathematica continually reinforces this preference for calculation over other possibilities like mapping the mathematics to the physical system at hand or packaging and evaluating mathematical chunks.

Did S1 realize she had to calculate something and then reach for Mathematica? Or did the chain start the other way, with Mathematica being within her reach, causing her to look preferentially towards calculation? Given the place we chose to start providing transcript, the former perhaps seems the most likely. However, S1 had promptly announced she had brought Mathematica with her back when she entered the room ten minutes earlier. Then again, maybe she had been vaguely aware of the tendency of quantum problems to involve calculation when she was packing her bag back at home that morning. It's a chicken and egg dilemma that we are not interested in teasing apart.

The important theme of this case study is that Mathematica is an active participant in how these students continually interpret and reinterpret their physics work, not merely a passive tool that offers them a convenient way to do whatever calculations they would have encountered on their own. This stickiness Mathematica gives the calculation preference is a significant source of difficulty in and of itself, even when a robust math sense is present in student thought.

## V. Connection to Cognitive Modeling

We have interpreted our results in terms of "sticky mindsets" facilitated by interaction with the symbolic manipulator; that is, students' cognition appear to be temporarily trapped in a local coherence of thought that inhibits direct and easy access to mental resources that would be relevant and useful. Phenomena of this sort are well known throughout the cognitive and behavioral sciences and affect much of everyday human behavior. We briefly review some of these results here. They fall under the rubrics of *framing* and *selective attention*.

Many academic disciplines have described how a person's existing knowledge and past experience affect present actions. All agree that humans do not approach situations as blank slates. As a quick example, consider entering a library. Even if you have never been in that particular library building before, you will immediately have a general idea how to proceed. You would expect there to be computers with easy access to the library's home catalog search page, stacks of books organized in a particular way, and copy machines. You would plan on doing certain types of work in this building like quiet reading, writing, and note taking. There would also be social expectations. You would not plan on shouting across a room or sprinting down an aisle.

Framing is probably the most common word for the, often subconscious, process by which our minds assess a new situation, bring relevant knowledge to bear, and suppress other knowledge deemed by the framing process to be irrelevant. Tannen[18] and MacLachlan and Reid[19] provide useful summaries of framing, as well as discussions of associated terms like schemas and scripts, across decades of research in a variety of academic fields including linguistics, psychology, sociology, anthropology, and art. All these researchers, however, are concerned with describing and analyzing the same general aspect of human cognition, what Tannen calls the "structure of expectations." Framing does not simply occur once at the beginning of a new situation; it is a process that is being continually updated. Perhaps the "library" sign in front of the building gives you a good idea of what to expect and how to act, but that judgment is continually reevaluated during your stay. If you see old-fashioned card catalogs rather than the computers you expect, you probably would still know how to find a book you wanted. Seeing other people behaving quietly feeds back into your own behavior.

We saw an example of this continual updating of the framing process in the roadblock analysis of S1's work. In lines 4, 9, 18, and 20-22, S1 is made explicitly aware that the calculation she had tried was not satisfactory. Her solution in each case was to try a different type of calculation. The behavior of others also influenced S1's thinking during the episode. When three other students offered more calculation suggestions in lines 15-17, those responses fed back into S1's thought and maintained her commitment to calculation.

Framing not only helps activate knowledge that one expects to be relevant in a given situation, it also inhibits knowledge that is seen as not currently relevant. When attending a play in a theatre, an usher quietly seating a latecomer may be comfortably ignored – and not even noticed.[20] When concentrating one's attention on a challenging task, such as counting the number of passes in a complex basketball event with more than one ball being exchanged, the viewer may miss noticing dramatic phenomena right in the center of the visual field.[21] In cognitive science, such phenomena are referred to as selective attention. They are often interpreted[22] as being a result of cognition having a limited amount of "attentional resources" that can be applied at any one time.[23]

Whenever a student works on a physics problem, he makes a judgment, usually implicitly, about what type of activity he or she is trying to do. This judgment primes certain resources and effectively inhibits the activation of other resources, leading the student to focus on a subset of their available knowledge while working.[24] This priming of a subset of one's resources is framing. Framing has also been discussed in the physics education literature.[24,25,26] Here we saw the framing process bracketing SA and SB's thought away from considering how their calculation techniques would relate to $<V>$. It also narrowed S1's search for reasons why her integral was diverging. She considered only reasons related to calculation instead of other possibilities like graphing the integrand or chunking the integrand and examining how $<x^2>$ and $\sin^2 x$ behave in the positive and negative infinity limits.

We have chosen to interpret our students' thinking in terms of framing and selective attention because we see these cognitive processes as the most fundamentally relevant ones and because this stance has been beneficial in past work.[24,25,26] We do not imply that other theoretical lenses are not applicable. Metacognition, for example, could almost certainly play a role in helping us understand these students' thinking. These students do not step back and ask, at least explicitly, questions like Schoenfeld's[27] "What exactly are you doing? Why are you doing it? How does it help you?" There is likely a connection between frequency of metacognitive events and flexibility of one's framing, but investigating such a claim is beyond the scope of this paper. Activity theory[28,29] advocates a much more socio-cultural analysis perspective. It holds that the use of tools, like Mathematica, is fundamentally a culturally determined process. Physics students tend towards calculation mindsets when they use Mathematica because that is how the tool is primarily treated among their social group. Again, we do not see this alternate perspective as orthogonal to the one we considered in detail. For conciseness, we have focused our analysis more at the individual cognitive level rather than the broader social level. A full treatment

## VI. Conclusions

We have spent the effort making a theoretical connection to framing for two reasons. First, it suggests a process by which Mathematica can help lead to the stickiness of the calculation tendencies that we have observed. This process, this framing the mind conducts, is much more general than some mental operation specific to a physics classroom alone. In a sense, it helps make the actions and shortcomings of the students in the earlier examples seem natural and reasonable. If our brains are indeed always involved in framing, always assessing situations, relating those assessments to groups of expectations, and allowing those expectations to limit our possible responses, then the temporary blindness these students exhibit becomes a plausible error.

Second, using framing to help model student thought, like any scientific theory, affects subsequent hypotheses we make about students' thinking. This influence is especially important in our real time interactions with students. As physics teachers, we will continue to encounter students using calculators and relying, at least in isolated episodes, too heavily on computation, as do the students in this paper. How should we address this issue in our classrooms? Being aware of these framing effects at least highlights an alternate cause of students' trouble, beyond simple inability. They may possess the relevant knowledge to solve their problem but are being actively bracketed away from this knowledge by their focus on their calculator. An appropriate response by the instructor might be to search for a trigger to this latent knowledge. We are not implying that all difficulties our students encounter can be adequately addressed by helping them reframe the issue. Sometimes there are gaps in their understanding and more direct instruction methods are appropriate. We are, however, arguing that framing issues are disproportionately often present when powerful calculators are involved because of the active role they assume in the dynamics of student thought.

In our interpretation of the events discussed here, Mathematica is an active participant in the students' framing of their approach. It provides feedback that encourages them to stay in a calculation mindset.

How can we justify this "active" interpretation? Basically all physics teachers have anecdotal stories of their students tending to prefer calculation over other modes of thought like mapping mathematics to the physical systems at hand. Perhaps these students were simply following this general trend, and it just happened that Mathematica was there. The strongest evidence against this passive interpretation comes from lines 15 to 18 in the second episode. These lines have three other members of the group, in addition to S1, chiming in with suggestions on how to resolve the infinite result problem. All of these suggestions refer to using Mathematica in different ways. At least at this particular time, Mathematica has become utterly ingrained in the students' thinking. Mathematica made them aware of the diverging integral problem in the first place, and all four of these resolution strategies involved using Mathematica again.

While we believe Mathematica is an active influence on students' thinking, we do not advocate its removal from the undergraduate physics curriculum. It certainly speeds up computation, and its graphing abilities can provide quick and detailed visualizations. Our purpose in presenting the analysis in this paper is twofold. As researchers, we wanted to argue for affording Mathematica and similar calculators the ability to drive students' thinking, often towards framing their activity as one of calculation. As teachers, we hope that detailing this phenomenon will make us more sensitive to its occurrence in our classrooms. If Mathematica is indeed an epistemologically potent tool, there is no reason not to explicitly address its power in class. Rather than simply suggesting students use Mathematica and leaving it to exert whatever influence it defaults to with each student, we could explicitly model specific uses of the program, using Mathematica to explore a function's behavior, to quickly test physically meaningful cases, to merely confirm a mathematical conclusion instead of generating it, and so forth. All the while, if we talk with students about how we are using the program in a way other than straightforward calculation, we may help them begin to see how to integrate their use of Mathematica with other mindsets that can broaden their math sense and make it more effective. Making students more explicitly aware of Mathematica's potential roles in their thinking is a first step to their learning to harness its full power for themselves.

## VII. Acknowledgments

The authors wish to thank the various members of the Physics Education Research Group at the University of Maryland for their many contributions to the analysis presented here. This work was supported by NSF grants DUE 05-24987 and REC 04-40113 and a Graduate Research Fellowship. Any opinions, findings, and conclusions or recommendations expressed in this publication are those of the author(s) and do not necessarily reflect the views of the National Science Foundation.